\documentclass[twocolumn,english,aps,prd,fleqn,floatfix]{revtex4}
\usepackage{epsf,amsmath}
\usepackage{graphicx}
\usepackage{rotating}

\usepackage[T1]{fontenc}
\usepackage[latin1]{inputenc}

\makeatletter
\usepackage{babel}
\makeatother
\begin{document}

\title{Novel ideas about emergent vacua}

\author{F.\:W.\:Bopp}
\email{bopp@physik.uni-siegen.de}
\affiliation{Department Physik, Universit\"at Siegen, D--57068 Siegen,
Germany}
\begin{abstract}
Arguments for special\emph{ emergent vacua} which generate fermion
and weak boson masses are outlined. Limitations and consequences of
the concept are discussed. If confirmed the Australian dipole would
give strong support to such a picture. Preliminary support from recent DZero
and CDF  data is discussed and 
predictions for LHC are presented.
\end{abstract}

\maketitle

\section{General Introduction}

The \emph{hierarchy problem} in particle physics is used as a guidance
to find theories beyond standard model~\cite{Ellis80}. The argumentation
in some way wrongly presumes a separation of particle physics and
cosmology. Without such a separation there is no need to directly
connect masses to Planck scale physics as a manageable scale is available
from the cosmological constant more or less corresponding to the neutrino
mass scale. As in Planck scale based models the spread in the fermion
mass scales will have to be explained. This is considerably easier
in a two scale model in which combinations of the GUT scale and dark energy 
scale can appear.

This argument seems  just to change the context as cosmology
contains a worse scale problem~\cite{weinberg89}: It is widely assumed that
the cosmological constant corresponds to the vacuum energy density caused
by a condensate~\cite{Bousso07,Bousso}. The properties of the condensate
have then somehow to reflect a Grand Unification Theory (GUT) scale
of the interactions when it presumably was formed (i.e. about $10^{15}$~GeV),
whereas the flatness of the universe requires a non-vanishing but
tiny cosmological constant (of about $3$~meV)~\cite{Frieman:2008sn}. 

The size of this gap rises a serious question. It is not excluded that
a dynamical solution of a type envisioned for the hierarchy problem
of particle physics~\cite{Susskind,Giudice:2007qj} will eventually
be found to be applicable. The opposite opinion seems more plausible. 
It considers it impossible to create such scales factors in a direct dynamical
way: A Lagrangian with GUT scale mass terms cannot contain minima
in its effective potential involving such tiny scales. Of course,
condensates do contain compensating energy terms, but without a new
scale true field theoretical minima have to stay close to the GUT
or Planck scale or they have to vanish.

Besides this plausibility argument there is a formal  problem with the conventional view. The term  hierarchy
problem is properly used if from a single available scale  
derived scales have to be obtained which are non-vanishing but  
many orders of magnitude away.
The discussed cosmological  problem is not a hierarchy problem. 
Other scales like the age of the universe are available which can bridge the
scale gap.

The observed tiny non-vanishing cosmological constant can just  means that
given the age of the universe the true minimal vanishing vacuum state is not reached~
\footnote{Studying the cosmological evolution similar ideas involving a time-dependent
vacuum state~\cite{Cai:2007us,Neupane,Krauss:2007rx} were considered.
However, based on the scale argument we here stick to the hypothesis
that the energy density of the true minimal state vanishes without
time dependence. An undisturbed vanishing vacuum energy also determining
the structure of the space was postulated in a less phenomenological
paper by~\cite{Volovik:2007fi,Klinkhamer:2009gm,Volovik:2010vx}.
Their basic premises exactly correspond in this point to discussed
concept. %
}. Our physical vacuum is then defined as an effective ground state. The
spontaneous symmetry breaking has to be replaced by an evolving process
yielding an unfinished vacuum structure. Its tiny mass density indicates
that the present condensate has to be quite close to a final one.
As vacuum-like state the condensate has to largely decouple from the
visible world. 

What could be the history of such a physical vacuum state? It is formed
at a condensate- (e.g. technicolor) force  mass scale, which
is assumed to more or less coincide which the GUT scale taken to be
somehow directly connect to the Planck mass. In a chaotic initial
phase bound composite states are formed with considerable statistical
fluctuations. As the mass scale and the potential scale are of the
same order these states can sometimes be more or less massless on
a  GUT scale. Their initial GUT scale geometric extend is  
not fixed dynamically by a mass scale. They or configurations
of them can reduce their remaining energy by geometrically extending.
In this evolving process they more and more decouple from the localized
hotter incoherent rest. This decoupling works only in one way. At their scale
they can, of course, radiate off energy eventually absorbed by the
much hotter rest. Cooling down eventually a quantum vacuum is formed
in a quantum mechanical condensation~\footnote{Such a quantum mechanical
condensation is reasonable. It is presumably actually  observed in the superfluid center of cold
neutron stars.}. Its properties have to be constant, 
perhaps almost over GLyr distances~\cite{webb_2010ki}.
The assumption is that it reaches coherence over a large, say galactic,  scale.
As explained below a lower limit of this coherency scale can be obtained from  weak interactions. 
Eventually if the universe would continue to expand it will of course reach 
a vanishing energy density~\footnote{Also the Casimir contribution to
the dark energy vanishes in this limit~\cite{Antoniadis:2006wq}}.

The advantage of this picture is that it requires no new scale. States
without scale are known from condensed matter physics and they are
called {}``gap-less''~\cite{Volovik:2007fi}. Without such an extra
scale the evolution of the dark energy in a comoving cell $\epsilon_{\mathrm{vac.}}$
has to be a linear decrease like:\[
\partial\epsilon_{\mathrm{vac.}}/\partial(\epsilon_{\mathrm{vac.}}t)=-\kappa\epsilon_{\mathrm{vac.}}\]
where $\kappa$ is a dimensionless decay constant and $t$ the time.
The absence of the usual exponential decrease has the consequence
that the age of the universe is no longer practically decoupling and
irrelevant. 

The expansion of the universe is not linear in time and in the above
equation the time $t$ has presumably to be replaced by the dynamical
relevant expansion
parameter $a$ to obtain a less rough estimate. It
yields $\epsilon_{vac.}\propto\frac{1}{a}$.
Phenomenologically
the expansion constant is taken to be $a\sim\sqrt{t}$ initially and
$a\sim t^{2/3}$ later on~\cite{Frieman:2008sn}. 
The ``hierarchy ratio'' of grand unification scale and present vacuum
scale \[
{\frac{\epsilon_{GUT}}{\epsilon_{vacuum}(t_{0})}=10^{27}}\]
can be obtained from the age of universe
\[t_{0}=5\cdot10^{46}\frac{1}{M_{GUT}}\]
with $a\propto t{_{0}}^{0.5\,\mathrm{resp.\,}0.66}$ to be \[
\frac{\epsilon_{GUT}}{\epsilon_{vacuum}(t_{0})}=2.2\cdot10^{23}\ \mathrm{resp.}\ 1.4\cdot10^{31}\,.\]
To the accuracy considered the problematic hierarchy ratio is explained. 

For the emergent vacuum reached in this way a rich and complicated
structure is natural. Heavier masses might involve combined scales.
As suggested by Zeldovich, Bjorken and others~\cite{Zeldovich,Bjorken:2010qx}
combinations of suitable powers of both scales: \[
H\, M_{\mathrm{Planck}}^{2}\sim\Lambda_{\mathrm{QCD}}^{3}\]
might explain needed intermediate mass values. The Hubble constant
$H$ is here related to the condensate mass density and $\Lambda_{\mathrm{QCD}}$
is the QCD mass scale. In chiral perturbation theory the mass of the
pion-like GUT-scale bound state can be estimated as~\cite{Leutwyler1996et}~:
\[ \begin{array}{ccl}
M^2 &=& B_\mathrm{condensate\ scale} \cdot \mathrm{(fermion\ mass\ scale)}=\\ 
 &=&(  3 {\mathrm meV} \cdot  10^{16}\, {\mathrm GeV} )^{\frac{1}{2}}= 170
{\mathrm GeV}
\end{array} \] All observed masses lie between the dark energy scale and
this Goldstone state scale.  

The above hierarchy argument is not new~\cite{BjTalk,Volovik:2010vx}.
Of course, the argumentation presented is rather vague. However, without
the constraint that the physical vacuum is at an actual minimum there
is too much freedom and it seems futile to try to obtain a more definite
description. A realistic ab initio description is presumably impossible.
Actually this lack is quite typical for most condensed matter in solid
state physics where the term Emergent Phenomena was coined for such
objects~\cite{Laughlin,Liu2005}~%
\footnote{As in \emph{geography} many properties of this \emph{Cosmographic Vacuum}~\cite{bopp:cosmo}
are dependent on a chaotic history and seemingly accidental. The name
of Cosmography was first used by Weinberg~\cite{weinberg}.%
}. 

Sometimes `emergent vacuum' is used synonym with the
'consecutively broken vacuum' of the standard model~\cite{Close}.
Here we take a narrower definition sometimes called strong emergence~\cite{Laughlin,LaughlinBook}
which includes basic unpredictability. The established complexity
of the known part of the vacuum legitimates this assumption. It has
two immediate consequences:
\begin{itemize}
\item[i)] It is extremely ugly from a model building point of view and actually\emph{
}leads to a murky situation: \emph{The vacuum is largely unpredictable
but predictions are necessary for the way science proceeds}. In this
way strong emergency is a physical basis of Smolin's wall~\cite{Smolin}
possibly severely limiting the knowledge obtainable. 
\end{itemize}
To proceed beyond this barrier is at least difficult. 
Quimbay and Morales~\cite{Quimbay}  assume an equilibrium and use the zero 
temperature limit of a thermal field theory. 
Volovik and Klinkhamer~\cite{Volovik:2010vx,Klinkhamer:2009gm}
try to rely on analogies to solid state physics. Bjorken argues~\cite{Bjorken:2010qx}
that the situation is somewhat analogous to the time around 1960 where
on had to turn to effective theories to parameterize the data. Here
we will not attempt to contribute to this difficult problem. 

The rich structure of the (strong) emergent vacuum has a second consequence:
\begin{itemize}\item[ii)]  The standard model contains many aspects with broken symmetries,
asymmetric situations and  partially valid conservation laws. In emergent vacuum picture many
of these observations might  not be truly fundamental and just reflect
asymmetries of the accidental vacuum structure. This takes away the
fundament of many theoretical considerations. Textbooks have to be worded 
more carefully.  
\end{itemize}
The second consequence of the emergent vacuum will be expanded 
here. Physics was often based on esthetic concepts. In this spirit we
postulate:  {\bf ``Fundamental physics should  be as simple as achievable.''}
With this postulate it is then possible
to come to a number of interesting consistency checks and testable consequences. 
The basic ignorance
of the vacuum keeps such predictions on a qualitative level. Formulated
in Bjorken's historic context today might be a time where simple minded,
Zweig-rule type phenomenological arguments~\cite{Zweig:2010jf} are
needed to sort things out. 

Section 2 will discusses general implications. Aspects connected with
fermion and boson masses follow in Sections~3 and 4. Section~5 turns
to indications from Fermilab and  predictions for LHC.

\section{General Consequences of Emergent Vacua}

One immediate outcome of this argument is the following cosmological
argument. Here no novelty is claimed and on the considered conceptual level it is 
contained p.e. in dark fluid models~\cite{Arbey}. In our context it is important 
as it invalidates an argument for unneeded new particles.   

As the present
vacuum is not at a unique point it has to be \emph{influenceable by gravity}~%
\footnote{The compressibility is here an effect of present day vacuum far away
from the Planck scale. We differ from Volovich et al.'s model~\cite{Volovik:2010vx}
in this point. His condensate stays on a Planck scale but decouples
from gravitons.%
}. The distinction between compressed dark energy and dark matter is
blurred\emph{.} Following the simplicity postulate we assume that
a suitable compressibility can eliminate the need of dark matter altogether
and lead to an effective MoND descriptio\emph{n}~\cite{Milgrom,Bekenstein:2010pt,Knebe:2009du}.
The changed power dependence predicted from the MoND theory for galactic
distances can be obtained if the extra compression-mass density of the condensate
drops off in the relevant region accordingly. There is no fine
tuning: All mass densities are more or less on the same order of magnitude.
Lorentz-invariance is no problem as the resulting effective theory
does not touch fundamental laws. We will see later how an almost
massless condensate mimics an effective relativistic invariance in
the world outside of the condensate. The offset between the centers
of baryonic and dark matter component seen after galaxy collisions~\cite{Bradac}
was said to contradict MoND theory. Here it constitutes no problem
as it takes cosmic times to rearrange the dark energy effects. 

Another important simple outcome is the not unique vacuum can act
as a \emph{reservoir}. It can have several consequences. 
We begin with the most drastic one, which would eliminate one of the most
ugly aspects in physics text books.
 
It is unsatisfactory and potentially problematic
to attribute the matter-antimatter asymmetry to the initial condition
of the universe. It is also widely agreed~\cite{Quinn} that no suitable,
sufficiently strong asymmetry generating process could be identified.
The emergent vacuum offers a simple way to abolish the asymmetry:
the vacuum can just contain the matching antimatter. The vacuum must be
charge-less and spin-less, but nothing forbids
it to contain a non-vanishing antifermionic density~%
\footnote{A recently proposed model of Hylogenesis~\cite{Davoudiasl} assumes
that the compensating antimatter sits in the dark matter part of the
vacuum, here not accepted as something separate. The
greek word means genesis of wood. It is used in philosophical texts as
general building material.%
}. 

Such a antifermionic density could be important for the stability
of the vacuum. Most known physical condensates are fermionic. As 
these extremely extended antifermionic states are
practically massless their Fermi repulsion will dominate. They provide
an anti-gravitating contribution in the cosmological expansion~%
\footnote{Models in which the pressure in the cosmological equation is a function
of the density were considered in ~\cite{Linder:2008ya}.%
} possibly replacing inflatons. Such a so called 'self-sustained' vacuum
was postulated by Volovik~\cite{Volovik:2010vx}. He assumes a filling with superfluid
$^{3}He\mathrm{-A}$ like atoms. Keeping the GUT scale condensation
force generic, the condensate can be partially characterized by its
color and flavor structure. We here assume that there is no lepton
asymmetry that a hadronic antifermionic structure suffices. There
are of course many possible contributions. In the following we consider
spin- and charge-less Cooper pairs of antineutron like states. 

We follow Bjorken who argues that the various components of the vacuum should not
be considered as separate objects~\cite{Bjorken:2010qx}. 
Our picture is
that the antibaryonic condensate, whose density is regulated
by Fermi repulsion, is accompanied by a somewhat less tidily bound
mesonic cloud largely responsible for the fermion masses. Both are
then seed to the known gluonic component mimicking the spontaneous
chiral symmetry breaking. 
Are there  problems with such a picture?

For such a vacuum structure there is a firm limit on the dielectric constant 
\[\frac{\epsilon(\mathrm{vacuum)}}{\epsilon_{0}}-1\approx\theta_{C}^{2}<10^{18}\]
from the unobserved vacuum Cherenkov radiation~\cite{Klinkhamer:2007ak}. 
However, the considered vacuum state is not excluded.
Its density is well known. As we see later the mesonic and the antibaryonic densities in such
a vacuum have to be of the same order of magnitude. The antibaryonic
density equals the baryon density outside of the vacuum which is:
\begin{eqnarray}
n_{\mathrm{baryon}} & =
\rho_{c}\cdot\Omega_{\mathrm{baryon}}/m_{\mathrm{neutron}} = \nonumber  & \\ 
  & =0.25\cdot\mathrm{m^{-3}}=1.9\cdot10^{-39}& (\frac{\mathrm{MeV}}{\hbar
c})^{3} \nonumber
\end{eqnarray}
The spin-less GUT scale bound states in the considered vacuum have no initial dipol moments.
A factor $(\,10^{18}\,\mathrm{MeV/\hbar c})^{-3}$ has to be added to
obtain the dielectric constant. The result
is well below the limit. 

The simplicity postulate presumably requires grand unification.
In a framework in which fermions represent a $SU[5]$, $SO[10]$ or
a similar gauge group, proton decay like processes  could present
a problem for the antibaryonic vacuum. Depending on the details of
the symmetry breaking and on evolution of temperature and fermion
masses processes like\[
\bar{d}_{right}\,\bar{d}_{right}\rightarrow d{}_{left}\,\nu\]
could occur without the protection of large mass scale differences
possibly destroying the  condensate. 

Presumably one can find a symmetry breaking and evolution path
which avoids the problem. A more 
drastic solution is the following: In the emergent-vacuum framework
left- and right-handed GUT partners do not have to correspond
to the mass partners~%
\footnote{The condensate binding would have to be generation dependent and in
analogy to the deuteron binding mixed generation components could be required in the condensate
(leaving the $nn$- or $pp$-state
unstable) .\\
As an example we  take   $SO(13) \to SO(10)\times SO(3)$ where the symmetry breaking is
assumed to involve $SU(5)$ and where the generational $SO(3)$ contains
only color triplets for fermion and gauge bosons.
Fermi\-statistics requires identical spin and isospin symmetry. For the
antineutron decay two $\bar d_L$  have to decay in a $q_R$ and a lepton.
As the right handed mass partner of the  $\bar d_L$ quarks will be in a
different  $SO(10)$ generation the produced quark will have a second
or third generation mass.\\
Decays involving neutrinos are then kinematically not possible.
Of course, decays involving charged leptons and  strange quarks could be possible. 
However they might have to involve $\tau$ leptons 
as the corresponding Cabibbo-Kobayashi-Maskawa matrix is unknown.
The required exact zero entries seem unnatural. However, the
evolution can  select stable vacuum and adjust the corresponding mass
matrix in a dynamical process.\\
In this framework the proton decay involves the same  $\tau$ lepton. }. %
Most decay channels of the antineutron-like states are then excluded. 
Two remaining channels involve a
charged lepton and a strange quark. It can be prevented by zeroes in the corresponding 
lepton-quark Cabibbo-Kobayashi-Maskawa matrix. The evolution of the vacuum could naturally select such a 
stable vacuum configuration. It would also  explain  the 
observed stability of the proton.

It is non-trivial to obtain a sufficiently uniform vacuum state. 
By themselves the vacuum states are extremely extended. Initial
statistical fluctuation could be augmented by magnetic effects in
a rapidly expanding universe~\cite{Sachs} separating different $U(1)$-charges
at leased for a time relevant for condensation.
Known condensation often involves replication processes which select
certain species and allow to amplify initial asymmetries over many
decades. Once an asymmetry between vacuum and visible world is established
annihilation processes within the vacuum radiating into the visible
world should purify its antimatter nature. 
The GUT scale condensation is thought to precedes at least part of
an inflationary period. In this way a relatively small area can be
magnified to extend over essentially our complete horizon~\cite{Linde:2005ht}.

There is, however, no reason that the tiny region we originate in
happens to have a constant (i.e. extremal) fermoinic density. In section
4 we will discuss how a higher antifermionic density increases the
vector boson mass. So we expect a temporal and spatial variation of masses.
As long as all masses vary in the same way it will be hard to observe.

The fine structure constant is not fundamental~\cite{webb_2010ki,Fritzsch:2009zz}. 
This holds  in any framework with a non-fundamental vacuum structure.  
The fine structure constant is determined
by $1/e^{2}=1/g^{2}+1/g'^{2}$ at the vector boson mass scale and
relates the fine structure constant to the in this context fundamental $U(1)$ and $SU(2)$
couplings.
As the renormalization scale dependence of the left hand and right hand
side is different the fine structure constant will depend on point
where the relation can be applied, i.e. on the non-fundamental $Z_0$ mass.
The fine structure constant deceases with increasing  $M_{Z_0}$.

As the fine structure constant  enters different optical spectral lines with 
distinct powers astronomical measurements in far away galaxies are
possible with high  precision. 
There is evidence for a spatial dependence of the form:\[
\Delta\alpha/\alpha\sim B\,\cos(\Theta)+m\]
where $B=1.1\pm0.8\cdot10^{-6}\,\mathrm{GLyr^{-1}}$, where
$m=-1.9\pm0.8\cdot10^{-6}$and where $\Theta$ is an angle 
to a specified sidereal direction~\cite{webb_2010ki}.
If confirmed this means that the expected spatial variation has a  GLyr scale. 
Naturally the antifermionic density and the vector-boson mass 
was higher in the past explaining
the observed reduction in the fine structure constant at $90^o$. 
The same sign in the variation of the fine structure constant
with time was indicated by the  LNE-SYRTE clock assemble preliminary 
yielding:
\[\frac{\partial}{\partial t}\alpha /\alpha = - 0.18 \pm 0.23 \ 
(10^{16}\mathrm{ 
year})^{-1} \:.\] 
involving  a different time period~\cite{Bize}.

The antimatter vacuum was introduced for reasons given above. It also
affects other symmetries. Whether the resulting consequences are consistent
offers a non trivial cross check.

\section{Vacuum and Fermion Mass Matrix}

How do fermions interact with the vacuum? As said the tiny energy
scale of the vacuum requires to a huge geometrical extension and the
momenta exchanged with the vacuum have to be practically zero. Such
interactions are described with scalar, first order terms of a low
energy effective theory~\cite{Leutwyler1996et}. Scalar interaction
with the (very light) vacuum state does not depend on its Lorentz system.
There is no contradiction to the observed \emph{Lorentz invariance}
in the outside world ~\cite{Klinkhamer:2008nr}. 

All masses have to arise from interaction with the vacuum. Their effective
couplings should be rather similar. The mass differences have to originate
in distinct densities of the components of the vacuum they couple
to. The excessive number of mass parameters is unacceptable for fundamental
physics. Here the problem is solved by attributing them to properties
of the emergent vacuum. The concept then conforms to Hawking's postulate~\cite{Hawking},
stating that `the various \emph{mass matrices cannot be determined
from first principles}'. The postulate doesn't preclude that certain
regularities might be identified and eventually explained~ \cite{Hall_Salem_Watari,Donoghue}. 

We denote vacuum-fermions with a subscript (\emph{V}). The relevant
interaction $q_{i}+(\bar{q}_{i})_{V}\to q_{j}+(\bar{q}_{j})_{V}$
in the lowest perturbative order is shown in Figure~1. Relying on
the Fierz transformation it can be shown to contain a scalar exchange.
In a theory without elementary scalar particles this flavor exchange
term is the only such contribution. %
\begin{figure}[h]
\begin{centering}
\includegraphics[scale=0.3]{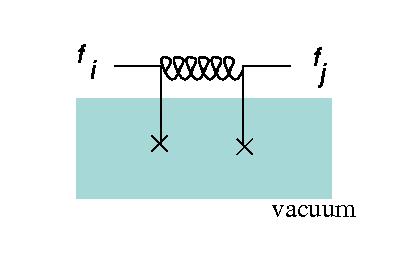}\vspace{-0.8cm}
\par\end{centering}

\caption{A process responsible for the fermion mass terms}

\end{figure}

We assume that such a flavor-dependent contribution stays important
if higher orders in the perturbative expansion are included. The matrix
elements then depend on the corresponding fermion densities and on
the properties of their binding, as interactions with fermions involve
replacement processes~\cite{Anderson,Quimbay}. Multi-quark baryonic
states should be more strongly bound than the mesonic states~\cite{Froggatt:2008hc}
and fermion masses should be dominated by the less tidily bound mesonic
contribution. In this way the required dominance of the mesonic $t\bar{t}$
contribution is not diluted by the light antiquarks from the antineutrons.

The\emph{ {}``flavor half-conservation''} is a serious problem in
the conventional view. Here the flavor of $q_{i}$ does not have to
equal the flavor of $q_{j}$. In this way flavor conservation\emph{
}can be restored and the apparent flavor changes in the visible world
can be attributed to a reservoir effect of the vacuum. As the vacuum
has to stay electrically neutral the mass matrix decomposes in four
$3\times3$ matrices which can be diagonalized and the CKM matrix
can be obtained in the usual way. If the coherent vacuum state is
properly included unitarity relations are not affected. As the matrix
elements mainly depend on fermion densities and not on messenger particles
with intermediate mass scales no significant scale dependence
is expected. In this way flavor changing neutral currents are suppressed
on a tree level. As in the standard model non tree level corrections
from weak vector mesons or heavier bosons are tiny. 

The antifermionic  vacuum state is obviously not symmetric under $CP$
and $CPT$ symmetry. This allows to restore these symmetries for fundamental
physics. Without any assumptions about discrete symmetries it is then
easy to see why $CPT$ is conserved separately in the outside world
and why $CP$ not. 

In the low momentum limit the interaction $f_{i}+(\bar{f}_{i})_{V}\to f_{j}+(\bar{f}_{j})_{V}$
will equal $\bar{f_{j}}+(\bar{f}_{i})_{V}\to\bar{f}_{i}+(\bar{f}_{j})_{V}$
by continuity in the exchanged momentum. In consequence the asymmetry
of the vacuum cannot be seen and $CPT$ is separately conserved in
the outside world. 

On the other hand the $(\bar{q}_{i})_{V}/(q_{i})_{V}$ asymmetry in
the vacuum will differentiate between $q_{i}+(\bar{q}_{i})_{V}\to q_{j}+(\bar{q}_{j})_{V}$
and $\bar{q}_{i}+(q_{i})_{V}\to\bar{q}_{j}+(q_{j})_{V}$ . In consequence
$CP$ appears as not conserved~%
\footnote{Consider the $(K_{0},\bar{K_{0}})$ system as an example. We assume
that such a pair was produced and that the $K_{0}$ remnant is observed
in the $(K_{0}^{L},K_{0}^{S})$ interference region. As postulated
above there are more $\bar{d}$ anti-quarks than $d$ quarks in the
vacuum. The amplitude, in which a pair of $d$ quarks is effectively
deposited in the vacuum during a $K_{0}\to\bar{K_{0}}$ transition
and then taken back later on during two $\bar{K_{0}}$ decays, obtains
a different phase as that of conjugate case of a pair of $\bar{d}$
anti-quarks deposited for the corresponding time. This changing phase
exactly corresponds to what is experimentally observed in $CP$ violation. %
}.%
\begin{figure}[h]
\begin{centering}
\includegraphics[scale=0.3]{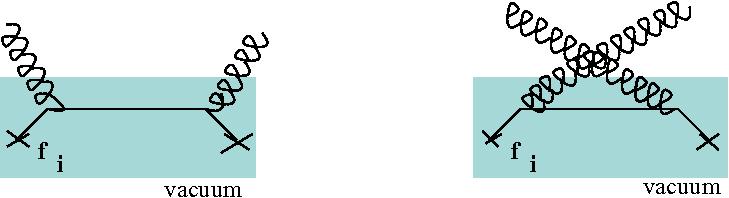}
\par\end{centering}

\caption{A process responsible for the vector boson mass terms}

\end{figure}

\section{Vacuum and Vector Boson Masses}

How do weak vector bosons obtain their masses? Relevant is a Compton
scattering process like $W_{\mu}+(\{\bar{q}\cdots\}_{i})_{V}\to W_{\nu}+(\{\bar{q}\cdots\}_{j})_{V}$
shown in Figure~2. 

In this framework pure gluon and meson condensates 
(p.e. simple Technicolor~\cite{Raby}
models with GUT binding scale) have a problem. 
In the low momentum limit the interaction
with the $B$-meson measures the squared charges of the vacuum content.
As these objects are $U(1)_{B}$ neutral states they cannot contribute
to a $m_{B}$-mass term. The appearance of antibaryonic states in
the vacuum provides a $U(1)_{B}$ charge. This has to  be considered as
an independent success of the antifermionic vacuum concept. 

Depending on the isospin structure of the bound states in the vacuum
the $m_{W}$-mass term can obtain contributions from the mesonic and
antifermionic component. It creates a $(B,W_{0})$ mass matrix, which
 diagonalizes in the usual way. The symmetry of this matrix allows
diagonalization and the electrical neutrality of the vacuum ensures
$m_{\gamma}=0$. 

One obvious question
is why the running Weinberg angle obtained from the  running
fundamental charges $g'$ and  $g$  equals a diagonalization angle
of the mass matrix. The question is related to the origin of Gell-Mann -
Nishijima relation $Q=T_{3}+\frac{1}{2}Y$.
Here the diagonalization angle has to take a value for which the antineutrons
of the vacuum are neutral at the symmetry breaking scale.

\section{Predictions for LHC}

Can one make predictions for LHC? Three {}``vacuum'' fluctuations
in bosonic densities are of course needed for the third component
of the weak vector bosons. The forces controlling these fluctuations must
allow for charge transfers analogously to  pion fields in nuclei.  Their
contribution will lead to
an effective scalar interaction with their longitudinal part shown
in Figure~3 manufacturing the mass and adding the third component~\cite{Zakharov:1997zt}. 

The rich flavor structure of the vacuum allows to excite many 
different oscillation modes.
Such phononic excitations
$H_{f_{i}\bar{f}_{j}}$ directly couple
to matching   $f_{i}\bar{f}_{j}$ pairs. Of course, three boson couplings
like $WWH_{f_{i}\bar{f}_{j}}$ can also appear. Their masses
should correspond within orders of magnitude to the weak vector boson
masses. 

Unlike GUT-scale bound objects their masses have to somehow reflect
the tumbled down vacuum scale. As described above
with GUT-scale constituent masses and with the tumbled down physical
condensate mass such pion-like states  could reach the TeV range.

\begin{figure}
\centering{}\includegraphics[scale=0.22]{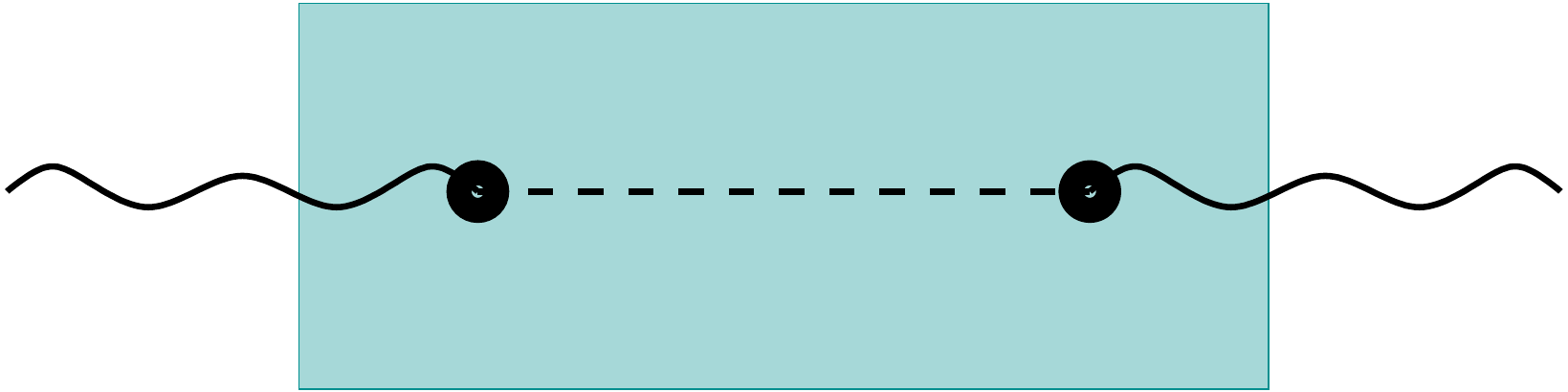}\caption{Mixing with the effective scalar}

\end{figure}

It is not difficult to distinguish such phonons from the usual Higgs
bosons~\cite{Higgs} as they couple to the fermions in a specific
way. In literature they are called {}``private Higgs'' particles~\cite{Porto},
which couple   predominantly
to one fermion pair. 

Their respective masses could reflect the constituents they are made of. 
The  light lepton and light quark phonons then have the lowest mass. 
The signal of a $H_{\nu \bar \nu }$ could be that of an
invisible Higgs boson~\cite{ZhuS.-h}. 
The absence of abnormal backward scattering in
$e^{+}e^{-}$annihilation at LEP could limit the corresponding leptonic-{}``Higgs''-boson to $M(H_{e^{+}e^{-}})>189\,{\rm
{GeV}\,}$\cite{Bhabha}.
Also the large-transverse-momentum jet production at Fermilab could limit 
the mass of light quark $H_{q \bar{q}}$
to an energy above $1{\rm \,{TeV}\,}$\cite{Fermilab}. 

However, the fermionic coupling constants are presumably much too tiny. 
The couplings to weak bosons and fermions are
drawn in Figure 4. In the limit of vanishing phonon momenta 
their couplings  correspond to the usual fermion mass terms. 
For heavier phonons this limit might be far away and meaningless. But 
for light phonons their  couplings should depend
on mass of the fermions involved similar to the usual Higgs bosons.
Then there is  no chance to observe the light phonons 
in fermionic channels. Their couplings to weak bosons is not
known. Very light bosons would be rather stable. With masses in the range of 
weak vector bosons  they could appear as quasi fermi-phobic {}``Higgs'' particles. 

For the very heavy quarks there is no problem with the coupling.
The $t\bar{t}$ -phonons will look like a normal Higgs 
allowing $t\bar{t}$ decays and will be hard to distinguish.
However, the expected mass range is different and such phonons 
are possibly out of reach kinematically.

The best bet might be the intermediate range. Here Fermilab collaborations
presented two canditates.

The first candidate is a $\tau \bar{\tau}$ -phonon in a mass range
 around $360$ GeV. It is seen as 
broad \{$e^{\pm}\mu^{\mp}$ missing $p_{\bot}$\}-structure in 
preliminary  D0 data~\cite{D0_Kraus}.
It is said to be statistically on discovery
level. At the moment D0 does not trust their $\mu$-energy calibration
sufficiently to announce it as such.

More significant is the $3.2$ sigma excess at $~150$ GeV in the dijet mass
spectrum of W + jets published by CDF~\cite{Aaltonen:2011mk}. One of the
virtual vector-boson decays  
$W^*\to W H_{s\bar{s}}$, $W^*\to W H_{c\bar{c}}$  or  $Z^*\to W H_{c\bar{s}}$ could
contribute in the observed region~\cite{Eichten:2011sh} in a way  comparable to the
seen processes
$W^*\to WZ$ or $Z^*\to WW$
yielding the observed two jet final state.

\begin{figure}
\centering{}\vspace*{-0mm}\includegraphics[scale=0.37]{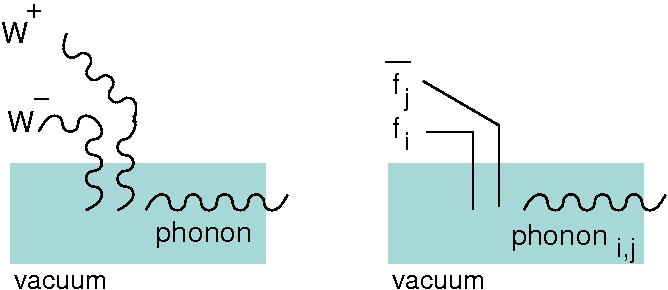}\caption{Coupling to phononic excitations}

\end{figure}

\section*{Conclusion}

The Emergent vacuum concept is not a beautiful scenario.
If correct the degree to which theory can be developed is quite limited
and one can forget the dream about the \emph{`Theory of Everything'}
at least as far as masses are concerned. However, the presented concept is not unpersuasive and things
fit together in a surprising way on a qualitative level. Firmly 
established private Higgs
particles would be an indication that an emergent scenario would be
\emph{nature's choice}.

\end{document}